\documentclass[amssymb,prb,floats,aps,superscriptaddress,twocolumn,narrowtext]{revtex4}
\usepackage{amsmath}
\usepackage{exscale}
\usepackage[mathscr]{eucal}
\usepackage{epsfig}

\begin{document}

\title{Superlight bipolarons and criterion of BCS-BEC crossover in cuprates}

\author{A. S. Alexandrov}
\affiliation{Department of Physics, Loughborough University,
Loughborough LE11 3TU, United Kingdom}

\begin{abstract}

Most of the proposed models of high-temperature superconductivity (HTSC) are
based on the short-range electron-electron correlations or/and on a
short-range electron-phonon interaction. However, in the cuprates the
screening is poor due to a low carrier density, layered crystal
structure, and high ionicity of the lattice. We develop further the
bipolaron model of HTSC, which explicitly takes into account the
long-range origin of all interactions. The long-range
electron-phonon (Froehlich) interaction binds carriers into real space
pairs (small bipolarons) with surprisingly low mass but sufficient
binding energy, while the long-range Coulomb repulsion keeps them from
forming larger clusters.  The model has explained many key
features of cuprates.  It is shown here that real-space pairing takes place in
many cuprates at variance with some (incorrect) criteria of the
BCS-BEC crossover.
\end{abstract}

\vskip2pc

\narrowtext

\bigskip
\maketitle

\section{Introduction: bipolaron model of cuprates}    
The electron-phonon  interaction as an origin of high $T_c$ continues to gather support through isotope
effect measurements, infrared, tunnelling, neutron and photoemission
spectroscopies \cite{sup}. To account for high values of T$_{c}$ in the
cuprates, it is necessary to have a strong electron-phonon
interaction, when a
many-electron-phonon system collapses into the small (bi)polaron
regime\cite{ALEX}. At first sight these carriers have a mass too large to be
mobile, however it has been shown that the inclusion of the on-site Coulomb
repulsion leads to the favoured binding of intersite oxygen holes\cite
{ALEXAND}. The intersite bipolarons can then tunnel with an effective
mass of only 10 electron masses\cite{ALEXAND,tru,alekor}

Mott and the author proposed a simple model\cite{MOTT} of the
cuprates based on bipolarons, Fig.1. In this model all the holes (polarons)
are bound into small intersite singlet and triplet bipolarons at any
temperature. Above $T_{c}$ this Bose gas is non-degenerate and below $T_{c}$
phase coherence (ODLRO) of the preformed bosons sets in, followed by
superfluidity of the charged carriers. The
model accounts for the crossover regime at $T^{\ast }$ and normal state
pseudogaps in cuprates\cite{mickab}. 
Here I show that real-space pairing takes place in
many cuprates.

\section{Pairing is `individual' in many cuprates}

The possibility of real-space pairing, as opposed to BCS-like pairing, has
been the subject of much discussion. Experimental and theoretical evidence
for an exceptionally strong electron-phonon interaction in high temperature
superconductors is now so overwhelming, that even advocates of nonphononic
mechanisms \cite{kiv} accept this fact. Nevertheless the same authors \cite
{kiv} dismiss any real-space pairing claiming that pairing is
collective in cuprates. They believe in a large Fermi surface with the
number of holes $(1+x)$ rather than $x$, where $%
x $ is the doping level like in $La_{2-x}Sr_{x}CuO_{4}.$ As an alternative
to a three-dimensional BEC of bipolarons these
authors suggest a collective pairing (i.e the{\it \ Cooper }pairs in the
momentum space) at some temperature $T^{\ast }>T_{c},$ but without phase
ordering. In this concept $T_{c}$ is determined by the superfluid density, which is
proportional to doping $x$, rather then to the
total density of carriers ($1+x)$. On the experimental side a large Fermi
surface is clearly incompatible with a great number of thermodynamic,
magnetic, and kinetic measurements, which show that only holes {\it doped }%
into a parent insulator are carriers in the {\it normal} state. On the
theoretical ground this preformed {\it Cooper}-pair (or phase-fluctuation)
scenario contradicts a theorem \cite{pop}, which proves that the number of
supercarriers (at $T=0$) and normal-state carriers is the same in any {\it %
clean} superfluid.

\begin{figure}[h]
\centering \epsfig{file=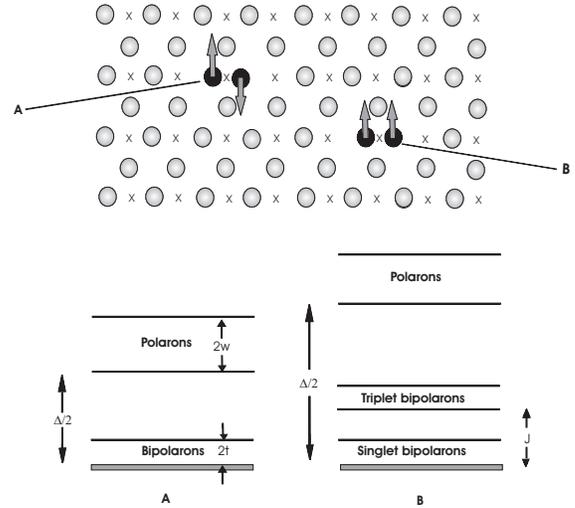, bbllx=50,
bblly=308, bburx=561, bbury=702,width=3.5in} \caption{Bipolaron
picture of high temperature superconductors. $A$ corresponds to
the singlet intersite bipolaron. $B$ is the triplet intersite
bipolaron, which naturally includes the addition of an extra
excitation band. The crosses are copper sites and the circles are
oxygen sites. w is the half bandwidth of the polaron band, t is
the half bandwidth of the bipolaron band, $\Delta/2$ is the
bipolaron binding energy per polaron and J is the exchange energy
per bipolaron.} \label{fig.1}
\end{figure}

Here I show that objections against real-space pairing also contradict a parameter-free
estimate of the Fermi energy \cite{aleF}. In cuprates the band structure is
quasi-two-dimensional (2D) with a few degenerate hole pockets. Applying the
parabolic approximation for the band dispersion we obtain the {\it %
renormalized} Fermi energy as 
$\epsilon _{F}=\hbar ^{2}\pi n_{i}d/(m_{i}^{\ast })$,
where $d$ is the interplane distance, and $n_{i},m_{i}^{\ast }$ are the
density of holes and their effective mass in each of the hole subbands $i$
renormalized by the electron-phonon (and by any other) interaction. One can
express the renormalized band-structure parameters through the in-plane
magnetic-field penetration depth at $T\approx 0$, measured experimentally: 
$\lambda _{H}^{-2}=4\pi e^{2}\sum_{i}n_{i}/(m_{i}^{\ast }c^{2})$. In
the framework of the BCS theory this
expression is applied to any $clean$ superfluid, where $m_{i}^{\ast }$
and $n_{i}$ are the $normal$ state mass and density of carriers, respectively. 
As a result, we obtain a {\em parameter-free } expression for the ``true''
Fermi energy as 
\begin{equation}
\epsilon _{F}={\frac{\hbar ^{2}dc^{2}}{{4ge^{2}\lambda _{H}^{2}}}},
\end{equation}
where $g$ is the degeneracy of the spectrum which might depend on doping in
cuprates. Because Eq.(1) does not contain any other band-structure
parameters, the estimate of $\epsilon _{F}$ using this equation does not
depend very much on the parabolic approximation for the band dispersion.

 Parameter-free estimates of the Fermi energy using Eq.(1)\cite{aleF} show that
the renormalised Fermi energy in many cuprates is certainly well below
$100$ $meV$ and the pairing 
is individual. Indeed such pairing will occur when the size of a pair, $%
\rho $ is smaller than the inter-pair separation, $r$. The size of a pair is
estimated as  $\rho =\hbar/(\sqrt{m^{\ast }\Delta})$,
where $\Delta $ is the binding energy. Separation of pairs is
directly related to the Fermi energy in 2D 
$r=\hbar \sqrt{\pi/(\epsilon _{F}m^{\ast })}$.
We see that the true condition for real-space pairing is 
\begin{equation}
\epsilon _{F}\lesssim \pi \Delta
\end{equation}

The bipolaron binding energy is thought to be twice the pseudogap. 
Experimentally measured pseudogap of many cuprates \cite{mickab}
is as large as $\Delta /2\gtrsim 50meV,$ so that Eq.(2) is well satisfied in
underdoped and even in a few optimally and overdoped cuprates. One should
notice that the coherence length in the charged Bose gas has nothing to do
with the size of the boson as erroneously assumed by some athors \cite{kiv}.
It depends on the interparticle distance and the mean-free path, and might
be as large as in the BCS superconductor. Hence, it is incorrect to
apply the ratio of the coherence length to the inter-carrier distance as a
criterion of the BCS-Bose liquid crossover. The correct criterion is given
by Eq.(2).


\begin{references}

\bibitem{sup} For recent results see  G. Zhao, M.B. Hunt, H. Keller and K.A. Muller, Nature, {\bf %
385}, 236 (1997); A. Lanzara, P.V. Bogdanov, X.J. Zhou, S.A. Kellar, D.L.
Feng, E.D. Lu, T. Yoshida, H. Eisaki, A. Fujimori,K. Kishio, J.I. Shimoyana,
T. Noda, S. Uchida, Z. Hussain and Z.X. Shen, Nature, {\bf 412}, 510 (2001)

\bibitem{ALEX}  A.S. Alexandrov, Phys. Rev. B, {\bf 46}, 2838 (1992).

\bibitem{ALEXAND}  A.S. Alexandrov, Phys. Rev. B, {\bf 53}, 2863 (1996).

\bibitem{tru}  J. Bonca J and S.A. Trugman, Phys. Rev. B {\bf 64}, 094507
(2001).

\bibitem{alekor}  A.S. Alexandrov and P.E. Kornilovitch, J. Phys. :Condensed
Matter {\bf 14}, 5337 (2002).


\bibitem{MOTT}  A.S. Alexandrov and N.F. Mott, J. Supercond (US), {\bf 7},
599 (1994).


\bibitem{mickab}  D. Mihailovic, V.V. Kabanov, K. Zagar, and J. Demsar,
Phys. Rev. B{\bf 60}, 6995 (1999) and references therein.

\bibitem{kiv}  E. W. Carlson, V. J. Emery, S. A. Kivelson, and D. Orgad,
cond-mat/0206217 and references therein.

\bibitem{pop}  A. J. Leggett, Physica Fennica {\bf 8,} 125 (1973); J Stat.
Phys. {\bf 93}, 927 (1998);

\bibitem{aleF}  A.S. Alexandrov, Physica C {\bf 363}, 231 (2001).

\end{references}
\end{document}